\newcommand{\ncom}{\newcommand}
\ncom{\be}{\begin{equation}}
\ncom{\ee}{\end{equation}}
\ncom{\eq}[1]{Eq.~(\ref{#1})}
\ncom{\rf}[1]{Ref.~\cite{#1}}
\ncom{\fg}[1]{Fig.~\ref{#1}}
\ncom{\mpsi}{M_{\rm PS}}
\ncom{\torstar}{\mu^*}
\ncom{\bpl}{\beta_{\rm pl}}
\ncom{\bplc}{\beta_{\rm pl}^{\rm c}}
\ncom{\GeV}{{\rm GeV}}
\ncom{\MeV}{{\rm MeV}}
\ncom{\order}[1]{{\mathcal O}(#1) }
\ncom{\vect}[1]{{\bf #1}}
\ncom{\bearray}{\begin{eqnarray}}
\ncom{\eearray}{\end{eqnarray}}
\ncom{\pl}{{\rm pl}}
\ncom{\rt}{{\rm rt}}
\ncom{\pg}{{\rm pg}}
\ncom{\nl}{\nonumber \\}
\ncom{\ls}[1]{\mbox{$\frac{1}{3}$\,Re\,Tr}(1-#1)}
\ncom{\zt}{Z_3}
\ncom{\vev}[1]{\langle #1 \rangle}

\input epsf

\documentstyle[epsf,12pt]{article}

\begin{document}

\rightline{SMUHEP 96-05}
\rightline{hep-lat/9605041}
\vspace{.3cm}
\rightline{May 1996}

\vspace{1.5cm}

\centerline{\Large\bf The Deconfinement Transition on }
\vspace{.2cm}
\centerline{\Large\bf Coarse Lattices}

\vspace{1cm}

\centerline{\large D.~W.~Bliss,$^a$ K.~Hornbostel,$^b$ G.~P.~Lepage$^c$}
\vspace{.5cm}
\centerline{\small $^a$University of California, San Diego, CA 92093}
\centerline{\small $^b$Southern Methodist University, Dallas, TX 75275} 
\centerline{\small $^c$Newman Laboratory of Nuclear Studies,
            Cornell University, Ithaca, NY 14853 }

\vspace{2.0cm}

\begin{abstract}
We compute the critical temperature $T_c$ for the deconfinement
transition of pure QCD on coarse lattices, with $N_t = 2, 3, 4$,
and lattice spacings from .33 fm to .15 fm.
We employ a perturbatively improved gluon action designed 
to remove order $a^2$ and $\alpha_s a^2$ errors.    
We find that $T_c$ in units of the charmonium 1P--1S
splitting and the torelon mass is independent of $a$ to
within approximately 5\%. 
\\[.3cm]
PACS number(s): 12.38.Gc, 11.10.Wx, 12.38.Mh, 11.10.Gh, 05.70.Fh
\end{abstract}

\pagebreak

\section{Introduction}

Recent calculations using improved lattice actions have demonstrated  
in several systems that accuracies of several per cent
could be achieved with lattice spacings $a$ as large as half a fermi.  
Use of an improved gluon action designed to remove order $a^2$ and 
$\alpha_s a^2$ errors demonstrated rotational invariance for the static 
quark potential independent of spacing errors to within 5\% on lattices 
with $a = .4$~fm\,\cite{coarse}.                               
Heavy quarkonium simulations with a nonrelativistic 
action corrected for order $v^2$ relativistic effects 
and $a^2$ errors produced masses with systematic errors 
as low as 5~MeV\,\cite{charm_ups,coarse}.                              
Other applications have included glueballs\,\cite{colin},                              
$B$ mesons\,\cite{bmesons}, light fermions\,\cite{fermions,fiebig} and                              
QCD at high temperatures\,\cite{beinlich}.                             

The cost of lattice simulations typically increase as $1/a^6$; 
improving accuracy by decreasing $a$ rapidly becomes prohibitively 
expensive.  An attractive alternative is to add effective interactions 
to remove spacing errors directly in the lattice action, bringing it
closer to the continuum\,\cite{syman}.  The use of tadpole-improvement                              
for the link variables and renormalized lattice perturbation theory 
in the design of these interactions are essential ingredients in these 
recent successes\,\cite{lm}.  Because these corrections can be computed  
perturbatively, parameters need not be tuned, nor new ones introduced 
beyond those of continuum QCD\,\cite{perfect}.                             

The efficiency with which improved actions remove lattice errors
dramatically increases the range of calculations accessible to 
currently available resources.  In this work we employ an improved gluon 
action to compute the deconfinement temperature of QCD.  In addition to
its intrinsic importance as a fundamental nonperturbative prediction, 
it provides an excellent quantity with which to test the accuracy of 
this and other improvement schemes\,\cite{cella,degrand,boyd}.

\section{Improved Gluon Action}

We used the improved action of \rf{Cur83,lw,coarse},
\bearray
S[U] &=& \beta_\pl \sum_\pl \ls{U_\pl} \nl
&+& \beta_\rt \sum_\rt \ls{U_\rt} 
+ \beta_\pg \sum_\pg \ls{U_\pg}\; ,
\eearray
with $U_\pl$ the Wilson loop about a single plaquette, $U_\rt$ a loop
around a $1\times 2$ rectangle,
and $U_\pg$ a parallelogram with links running along the opposing
edges of a cube.  These terms are sufficient to provide the dimension     
four and six operators needed for an effective action accurate to 
order $a^4$.

The coefficient $\bpl$ is the only input parameter, while 
\be
\beta_\rt = -\frac{\beta_\pl}{20\,u_0^2}\, \left( 1 + 0.4805\,\alpha_s
    \right),  \;\;\;\;\;
\beta_\pg = -\frac{\beta_\pl}{u_0^2} \, 0.03325\,\alpha_s
\ee
were computed in terms of $\bpl$ using tadpole-improved perturbation
theory\,\cite{w}.  The division of each link by $u_0$ gives 
tadpole-improved operators.  These ensure that the connection 
between these quantities and their continuum equivalents are not 
ruined by large tadpole contributions, and are essential for the 
effectiveness of lattice perturbation theory and the improved 
action\,\cite{lm}.    The measured expectation value of the plaquette 
determines both $u_0$ and the renormalized coupling constant $\alpha_s$
self-consistently, with 
\be
 u_0 = \left(\mbox{$\frac{1}{3}$\,Re\,Tr}
   \langle U_\pl \rangle\right)^{1/4}, \;\;\;\;\;
 \alpha_s = -\frac{\ln\Bigl({\textstyle {1\over 3}}\mbox{Re\,Tr\,}
   \langle U_\pl\rangle\Bigr)}{3.06839}\; .
\ee

\section{Deconfinement Transition and \\Critical Temperature}

Green's functions computed on a Euclidean lattice periodic in time
with finite extent $t$ are equivalent to those at the finite
temperature $T = 1/t$, averaged with the usual Boltzmann weight
$\exp(-H/T)$.  
In particular, the correlation of a Polyakov loop or Wilson line,
\be
 P(\vect{x}) \equiv {\rm Tr} \prod_t^{N_t}{U_{(\vect{x},t),\hat{t}}} \; ,
\ee
which wraps around the lattice in the time
direction, and its adjoint are related to the free energy $F_{Q\bar Q}$
of a static quark-antiquark pair separated by $\vect{x}$ according to
\be
 \vev{P(\vect{x}) P^\dagger(0)}\;\; \propto\;\; \exp(-F_{Q\bar Q}/T) \; .
\ee
At large separation, cluster decomposition requires
\be
 \vev{P(\vect{x}) P^\dagger(0)} \longrightarrow |\vev{P(0)}|^2 \; .
\ee
In the confining phase, $F_{Q\bar Q}$ increases with $|\vect{x}|$
and $\vev{P(0)}$ must therefore vanish; in the deconfined phase,
$P$ may acquire a nonvanishing expectation value.  Pure SU(3)
lattice gauge theory, including improved actions, possesses a global 
symmetry under multiplication of timelike links at a single time slice 
by the elements of the group center, $\zt$.  Nonzero $\vev{P}$ indicates 
the breakdown of this symmetry. Therefore $\vev{P}$ is
a suitable and relatively simple order parameter with which to study
this transition\,\cite{decon}.   

As the temperature exceeds its critical value $T_c$, the system
moves from confined to deconfined phases.  Measurements of $P$,
which were clustering around zero, now cluster about the real axis,
or to the two axes along the other $\zt$ elements, $\exp(\pm i 2\pi/3)$.
During the transition, the system will fluctuate 
between phases.  We determine $T_c$ by the location of the peak in 
the susceptibility
\be
  \chi \equiv N_s^3 (\vev{P^2} - \vev{P}^2)
  \label{eq_chi}
\ee
which measures the deviation in $\vev{P}$ these fluctuations 
induce.  Here $P$ is averaged over spatial lattice sites.

Because the volume is finite, the transition temperature is smeared,
limiting the precision with which we may determine $T_c$.
Also, above $T_c$, given sufficient simulation time the system will 
tunnel between the three degenerate vacua, and $\vev{P}$ will still 
average to zero.  It is therefore useful in practice to fold 
these three axes together\,\cite{fukugita,cella}, 
replacing $P$ in \eq{eq_chi} with $\Omega$, such that
\be
  \Omega \equiv  \left\{
    \begin{array}{l l}
      Re \: P & ; \:  arg \: P \:
      \epsilon \: (-\pi/3, \pi/3) \cr
      Re \: P \: e^{i 2\pi/3}  & ; \: arg \: P \:
      \epsilon \: (\pi, -\pi/3)   \cr
      Re \: P \: e^{-i 2\pi/3}  & ; \:  arg \: P \: 
      \epsilon \: (\pi/3, \pi) \; .  \cr 
    \end{array} \right. 
  \label{eq_omega}
\ee

We move our lattice system through the phase transition by holding 
$N_t = 1/aT$ fixed while adjusting $\bpl$, and therefore implicitly 
the lattice spacing $a$.  We are measuring a quantity defined at 
infinite spatial volume, and $N_t$ should be small relative to the 
number of spatial sites per side, $N_s$.  We determined $\bplc$ on 
lattices with $(N_s^3 \times N_t)$ of $(4^3 \times 2)$,
$(6^3 \times 2)$, $(6^3 \times 3)$, $(9^3 \times 3)$ and 
$(8^3 \times 4)$.

After determining the critical coupling $\bplc$ at which the 
transition occurs, we computed a convenient mass $M$ with this
same $\bplc$ but at zero temperature, with $N_t > N_s$.
The value of $M a$ from this simulation 
fixes $a$, and therefore $T_c$, in units of $M$.  Comparison with
data for $M$ gives $T_c$ in physical units.  We chose 
for $M$ the charmonium $1P-1S$ splitting, and also the torelon
mass, which is related to the string tension.

To determine $\bplc$, we began with $\bpl$ below transition
and used one thousand Metropolis updates to allow the lattice
to thermalize.  We then sampled the Polyakov loop after every 
five updates for the next one or two thousand.  After stepping $\bpl$ up
and allowing several hundred more updates, we began sampling again,
repeating until $\bpl$ had crossed its critical value.
To ensure that the number of updates was sufficient to thermalize,
we stepped $\bpl$ back down through the transition to check for
hysteresis.

\begin{figure}[thbp]
  \begin{minipage}[t]{2.5in}
    \epsfxsize=2.5in
    \epsfbox{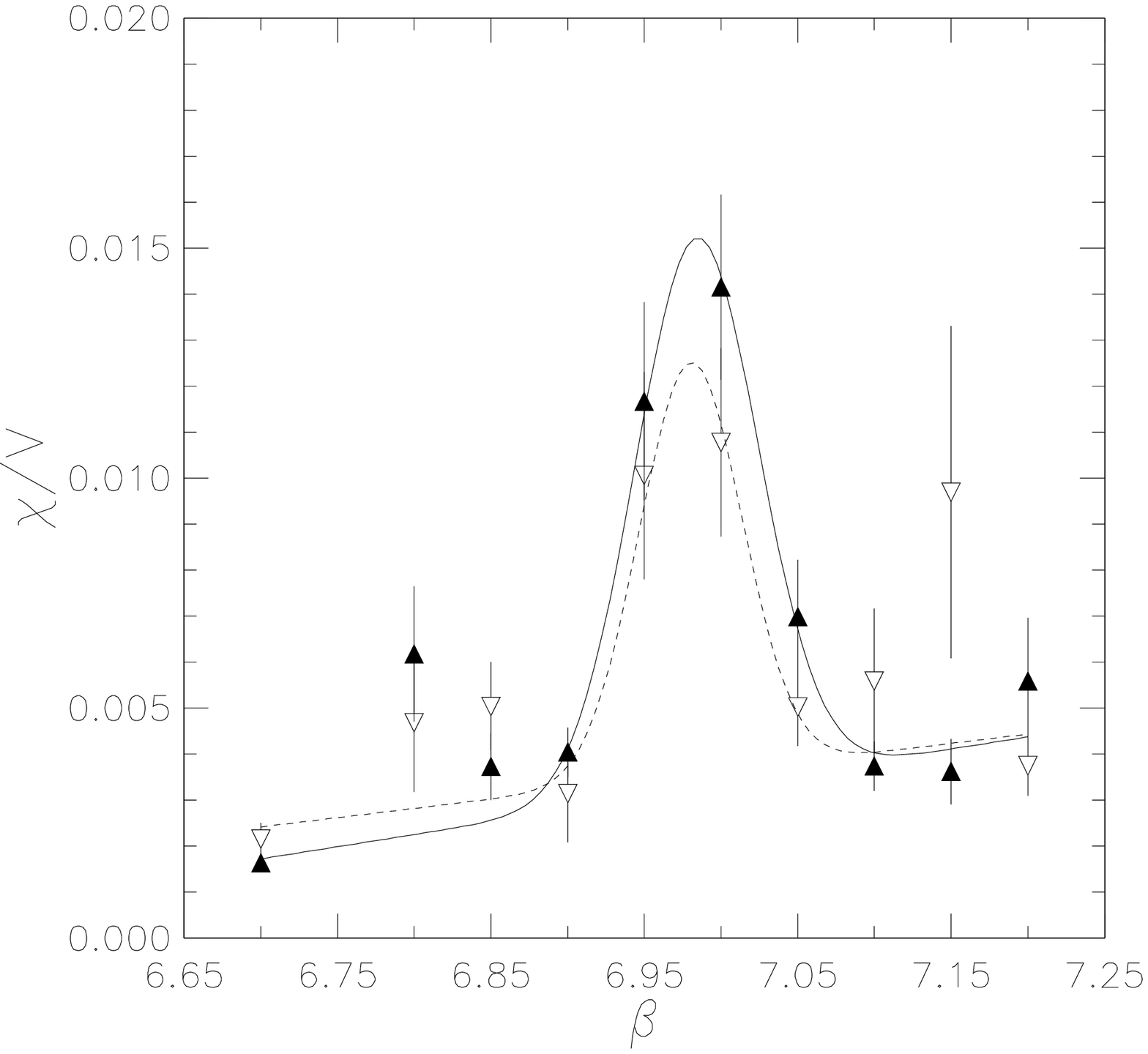}
    \caption{Susceptibility $\chi$ for \hbox{$4^3\times 2$} lattice.
      Arrows indicate upward or downward sweeps of $\bpl$
      through transition.}
    \label{fig_4x2_sus}
  \end{minipage}
  \hfill
  \begin{minipage}[t]{2.5in}
    \epsfxsize=2.5in
    \epsfbox{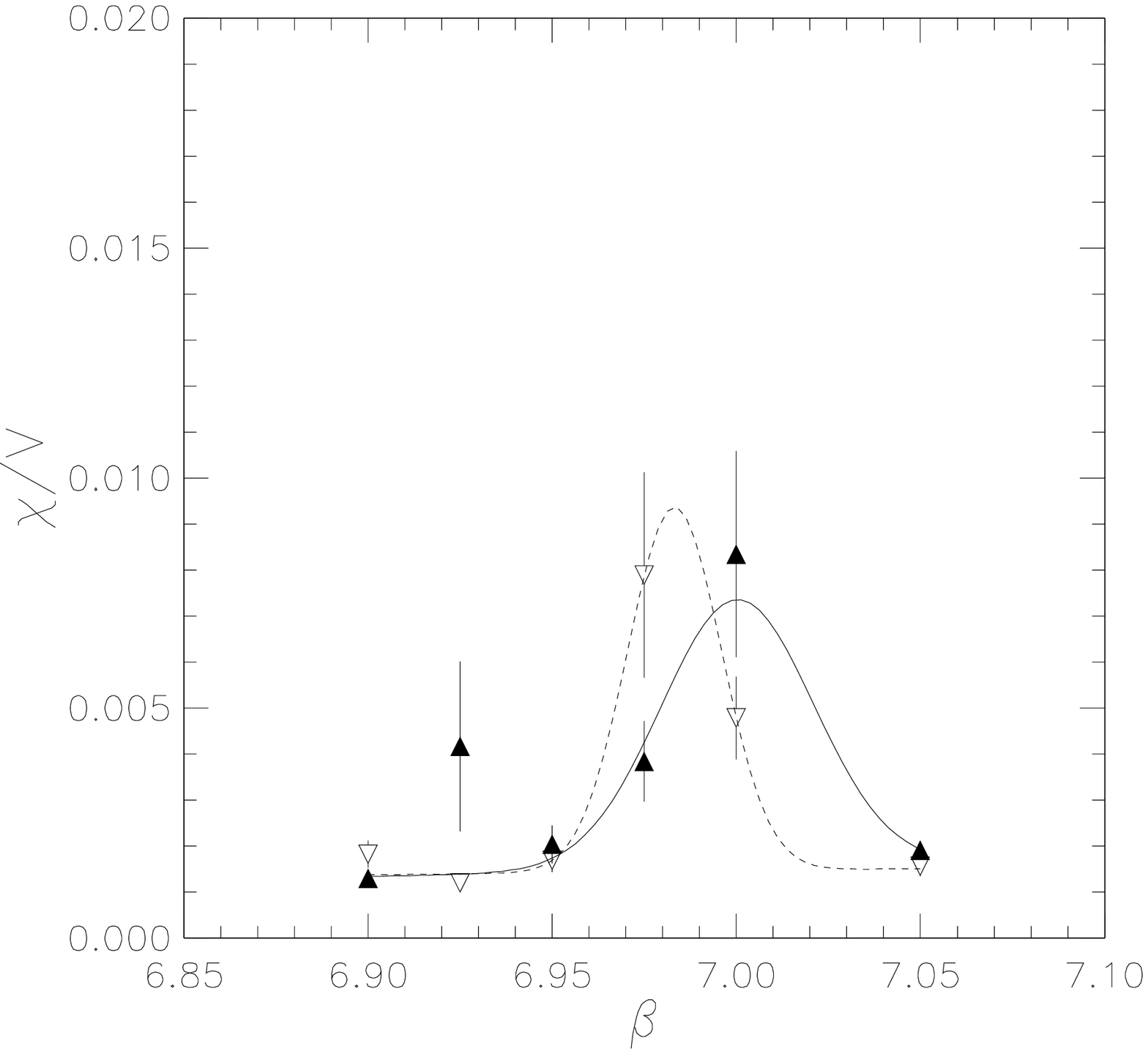}
    \caption{$\chi$ for $6^3 \times 2$ lattice.}
    \label{fig_6x2_sus}
  \end{minipage}
\end{figure}

\begin{figure}[thbp]
  \begin{minipage}[t]{2.5in}
    \epsfxsize=2.5in
    \epsfbox{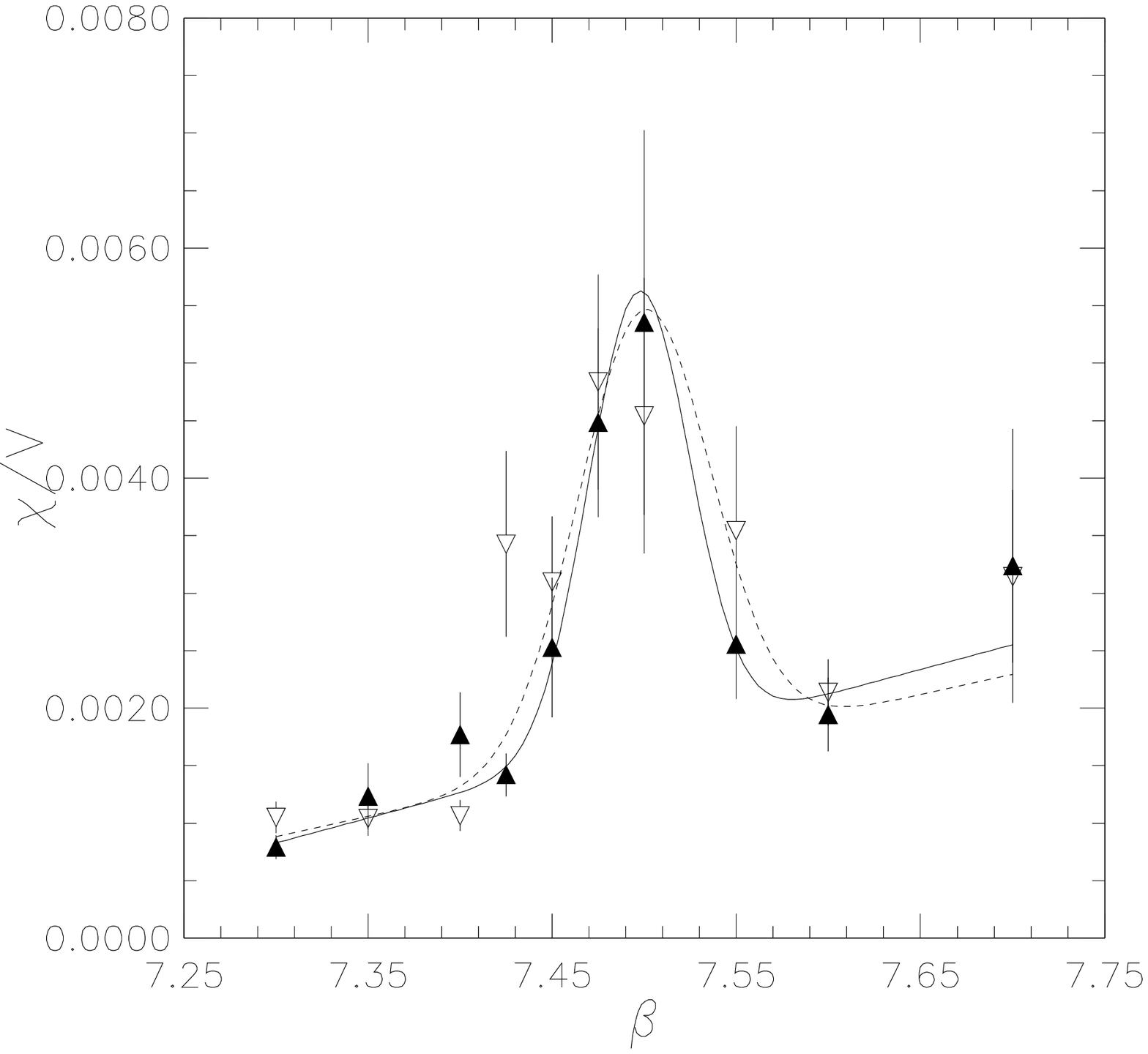}
    \caption{$\chi$ for $6^3 \times 3$ lattice.}
    \label{fig_6x3_sus}
  \end{minipage}
  \hfill
  \begin{minipage}[t]{2.5in}
    \epsfxsize=2.5in
    \epsfbox{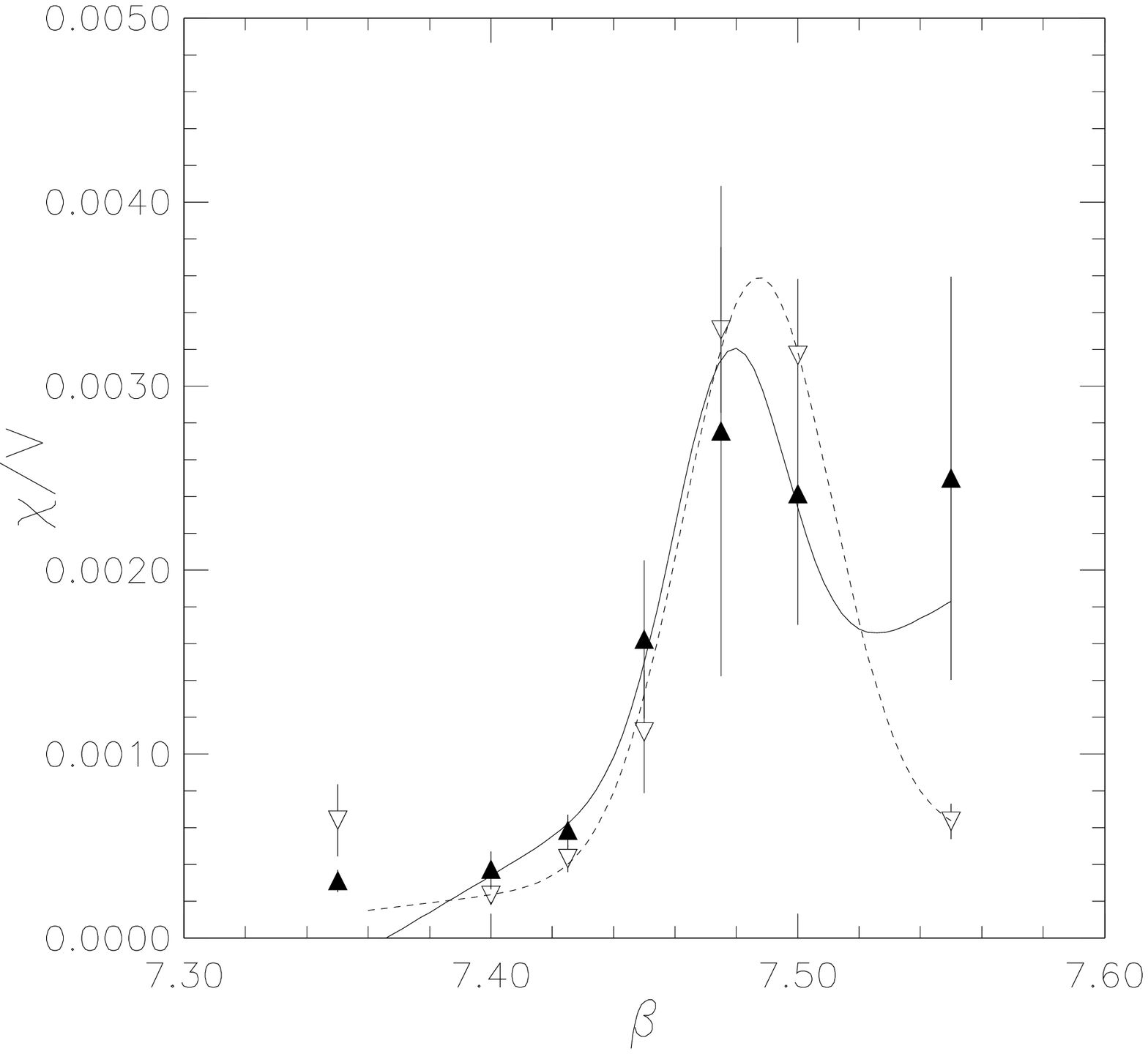}
    \caption{$\chi$ for $9^3 \times 3$ lattice.}
    \label{fig_9x3_sus}
  \end{minipage}
\end{figure}

\begin{figure}[thbp]
   \centering \leavevmode
   \epsfxsize=2.5in
   \epsfbox{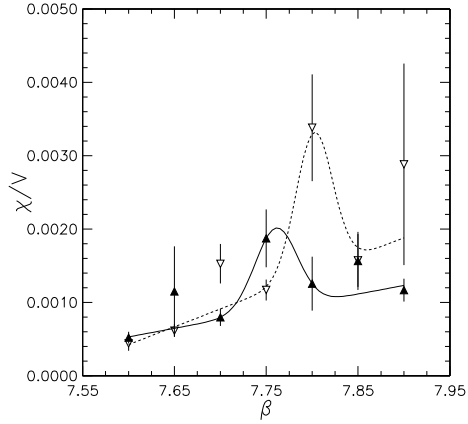}
    \caption{$\chi$ for $8^3 \times 4$ lattice.}
   \label{fig_8x4_sus}
\end{figure}

Figs.~\ref{fig_4x2_sus} through \ref{fig_8x4_sus} display the 
susceptibility plots for each lattice as a function of $\bpl$.
The up and down arrows indicate data from upward and downward
sweeps.  The solid and dashed lines show
independent fits of each to a Gaussian and a first order polynomial.  
The error bars indicate bootstrap estimates.  We determine $\bplc$ from 
the location of the peak, and assign the Gaussian width as the error.
Table~\ref{tab_crit_beta} presents the weighted average of 
$\bplc$ for each $T_c$. 


\begin{table}[thbp]
\centering
\begin{tabular} {|c|ccc|} \hline
$T_c$ & $1/2$ & $1/3$ & $1/4$ \\[.1cm] \hline
$\bplc$ & $6.99(1)$ & $7.49(1)$ & $7.78(2)$ \\[.1cm] \hline
\end{tabular}
\caption{Critical coupling $\bplc$ as a function of critical temperature.
$T_c$ is in lattice units.}
\label{tab_crit_beta}
\end{table}

While we found locating the peak in $\chi$ the most reliable method
for locating $\bplc$, an alternative is to locate the 
phase change by directly observing $P$ as a function of $\bpl$.
\rf{columbia} implements this by choosing a threshold in the 
fraction of $\vev{P}$'s found near any $\zt$ axis.  
\fg{fig_polyakov} displays the real and imaginary parts of $P$
measured during an upward sweep of $\bpl$ on the $9^3 \times 3$ lattice.
The transition is evident.  Below $\bpl$ of about 7.45, $P$ clusters
about the origin.  During the transition it explores the three
degenerate vacua which break $\zt$, before settling into one above
the transition.  Given sufficient simulation time, $P$ will populate 
each arm equally.  In our study, it rarely explored more than two
of the three.  

\begin{figure}[thbp]
   \centering \leavevmode
   \epsfxsize=5.0in
   \epsfbox{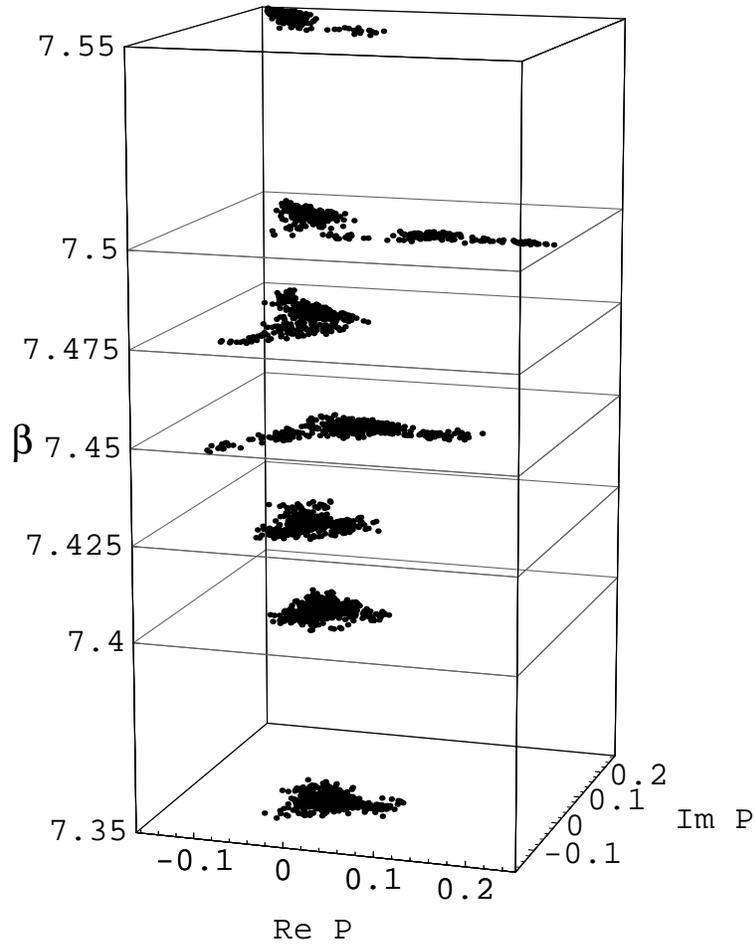}
   \caption{Real and imaginary parts of Polyakov loops measured
            on a $9^3\times 3$ lattice, during an upward sweep of
            $\bpl$ through $\bplc$. }
   \label{fig_polyakov}
\end{figure}

\section{Charmonium}

To test the dependence of these results on the lattice spacing and to 
convert the critical temperatures to physical units, we computed the 
mass difference $\mpsi$ between the $1P$ and $1S$ states of charmonium 
on lattices with the same spacings at zero temperature.  We chose this 
quantity because it is experimentally well-measured, simple to compute 
accurately using a nonrelativistic action 
(NRQCD)\,\cite{improved,charm_ups}, small and therefore 
insensitive to finite lattice errors, and its splittings are essentially 
independent of the charm quark mass.  We used an NRQCD quark action which
removes the leading order $a^2$ and $v^2$ errors in lattice spacing and 
quark velocity, and so is consistent with the gluon action.  The dominant 
uncorrected error is $\order{a^2 v^2}$, due to the $a^2$ error in the 
lattice version of the $\vect{D}\cdot\vect{E}$ relativistic 
$\order{v^2}$ correction.  We did not include dynamical light quarks.  
Spin splittings are absent to this order, and we compare our results 
to spin-averaged data.

\begin{table}
\begin{center}
\begin{tabular}{cccccc}
\hline
$\bplc$ & $T_c$ & lattice & $M_C$ & $\mpsi$ & $a^{-1}$~[GeV] \\ \hline
 6.99  & 1/2 & $3^3\times 12$ & 2.0 & .77(1) & .596(8) \\
 6.99  & 1/2 & $4^3\times 12$ & 2.0 & .794(4) & .577(3) \\
 7.49  & 1/3 & $6^3\times 12$ & 1.3 & .494(6) & .93(1) \\
 7.78  & 1/4 & $6^3\times 12$ & .98 & .378(7) & 1.21(2) \\
 7.78  & 1/4 & $8^3\times 12$ & .98 & .375(4) & 1.22(1) \\
\hline
\end{tabular}
\end{center}
\caption{$\mpsi$ as a function of $\bplc$.  Errors are statistical.
The critical temperature $T_c$, the charm mass $M_C$, and the
$1P-1S$ splitting $\mpsi$ are in lattice units.  The experimental
value for $\mpsi$ fixes $a$.}
\label{psi_table}
\end{table}

Table \ref{psi_table} presents the values for $\mpsi$ for these 
lattices.  We tuned the charm quark mass $M_C$ by requiring that the 
energy for states carrying nonzero momentum satisfy the usual 
nonrelativistic dispersion relation.  Comparing the lattice result to 
the experimental spin-averaged value for this splitting, 
$\mpsi = 457.9~\MeV$, fixes the inverse lattice spacing. 

We obtained these masses by $2\times 2$ correlated matrix fits of 
exponentials to three propagators for which either the source, 
the source and sink, or neither were smeared by the lattice 
equivalent of $\exp(-p^2/12) \sim (1 + \nabla^2/12)$, designed to 
suppress states with large momentum.  We fit to both the ground and 
first excited states in each channel.  Including the excited state 
prevented it and higher states from contaminating the value for the 
ground state, and allowed us to include data from lower times.  
We found this procedure useful in \rf{charm_ups}, and especially helpful 
in extraction of torelon masses, discussed below.  Typically we 
obtained good fits by including minimum times from 3 to 6, to maximum 
times from 9 to 12.

\begin{table}
\begin{center}
\begin{tabular}{ccccc}
\hline
$\bplc$ & $T_c$ & lattice & $T_c/\mpsi$  &  $T_c$~[MeV] \\ \hline
 6.99  & 1/2 & $3^3\times 12$ & .651(14) &  298(6) \\ 
 6.99  & 1/2 & $4^3\times 12$ & .630(10) &  288(5) \\
 7.49  & 1/3 & $6^3\times 12$ & .675(13) &  309(6) \\
 7.78  & 1/4 & $6^3\times 12$ & .661(17) &  303(8) \\
 7.78  & 1/4 & $8^3\times 12$ & .667(14) &  305(7) \\
\hline
\end{tabular}
\end{center}
\caption{
Critical temperature $T_c$ in units of $\mpsi$ and in MeV.}
\label{psi_scaling}
\end{table}

Table \ref{psi_scaling} displays the critical temperature in units of
$\mpsi$ and in MeV, using the experimental value for $\mpsi$. 
\fg{fig_beta_crit} presents the average $T_c$ for each $a$,
around the central value $T_c = 300(6)~{\rm MeV}$.  Along with the 
statistical errors of quoted for $\mpsi$ in Table~\ref{psi_table},
we include errors associated with uncertainties in the critical value for 
$\bpl$ and in the value for the average plaquette $u_0$, which is used to 
tadpole-improve the operators in the action which remove the $\order{a^2}$ 
errors.  To determine the former, we used one-loop perturbative
evolution to estimate the variation of $\mpsi$ with $\bpl$.  For the
latter, a variation in $u_0$ would affect the ability of the improved 
action to remove $\order{a^2}$ errors.  We estimated the percentage 
error by $\delta u_0/u_0$ times the typical (and similar) 
$\order{a^2}$ errors for these systems:  $a^2 \vev{p^2}$ for 
charmonium, and $a^2 \sigma^2$ for the torelon, discussed below.  
Here $\sigma$ is the string tension, and $\delta u_0$ 
the difference between the input and measured value.  We found
these methods to provide a reasonably reliable estimate 
of the variation of $\mpsi$ and the torelon mass
in the data over a fairly broad range in $\bpl$ and $u_0$.

The values for $T_c$ are nearly consistent within these errors, and
show little dependence on lattice spacing.  We note that values for 
$T_c/M_{PS}$ for the coarsest lattices ($\bplc = 6.99$) deviate by 
about two standard deviations.  This is a possible indication of 
the $\order{a^4}$ systematic errors for which the action is uncorrected.
These might be expected to range from about 5--10\% for the coarsest 
lattice (about 1/3 fm) to a couple per cent for the finest (1/6 fm).  

\begin{figure}[hbp]
   \centering \leavevmode
   \epsfxsize=2.5in
   \epsfbox{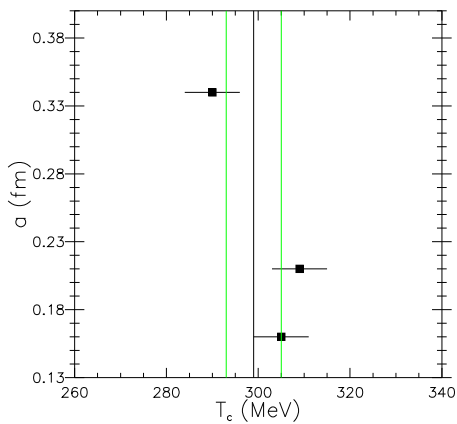}
   \caption{Critical temperature $T_c$ as a function
     of lattice spacing $a$, around central value.  
     $\mpsi$ sets the physical scales.}
   \label{fig_beta_crit}
\end{figure}

\section{Torelon Mass}

\begin{table}
\begin{center}
\begin{tabular}{ccccc}
\hline
$\bplc$ & $T_c$ & lattice &  $\mu$ & $\torstar$  \\ \hline
 6.99  & 1/2 & $3^3\times 12$ & 1.46(2) & --- \\
 6.99  & 1/2 & $4^3\times 12$ & 2.26(5) & --- \\
 7.49  & 1/3 & $6^3\times 12$ & 1.42(1) & 3.0(4) \\
 7.78  & 1/4 & $6^3\times 12$ & .676(6) & 1.8(3) \\
 7.78  & 1/4 & $8^3\times 12$ & 1.05(2) & 2.1(1) \\
\hline
\end{tabular}
\end{center}
\caption{Torelon mass $\mu$ and excited torelon mass $\torstar$ vs. 
$\bplc$. Errors are statistical.
$T_c$ is the critical temperature in lattice units.}
\label{tor_table}
\end{table}

As a second test of scaling, we measured the torelon mass $\mu$ for 
torelons of two different physical lengths at these same $\bpl$.  
The torelon is created by a Wilson loop strung 
along the full spatial extent of the periodic lattice, and
averaged over spatial orientations and transverse 
positions\,\cite{torelon}.  As such, it is sensitive to both the 
string tension $\sigma$ and lattice length $L$.  

We obtained the torelon mass from the exponential decay in the
correlation between two such loops at different times.  More precisely,
we accounted for periodicity in time by fitting to 
hyperbolic cosines rather than exponentials, 
and typically fit to time steps from one to six.  
For sufficient length $L$, the mass should go as 
\be
 \label{musigma}
 \mu \sim \sigma L\; , 
\ee
reflecting the cost in energy of keeping static charges separated by 
this distance.  We will use $\mu$ to determine a value for $\sigma$ 
below.

We may estimate the energy expected for excited states by comparing 
to those of a classical vibrating periodic string, $2\pi n/L$.  
This would suggest first excited states split from the ground states 
by $2\pi/N_s$ in lattice units, close to the values we observe.  
For larger lattices, it therefore becomes increasingly important to 
include these states in fits.  

To extract the torelon masses, we performed correlated $3\times 3$ 
matrix fits of three hyperbolic cosines to six torelon propagators 
which were smeared once, twice or unsmeared at either the source or sink.  
To smear these, we used the operator $\exp(-\alpha p_\perp^2)$ on the link 
variables\,\cite{ape}, which suppressed contributions with large momentum 
transverse to the loop.  Specifically, we applied $n$ times the lattice 
version of $(1 + \alpha \nabla_\perp^2/n)$, with $n$ typically 7 and 14, 
and $\alpha$ effectively .6 and 1.2.  We reuniterized the link variables 
between applications.  

\begin{table}
\begin{center}
\begin{tabular}{ccccccc}
\hline
$\bplc$ & $T_c$ & $L$ & $T_c/\mu$ & $T_c/\torstar$  & 
$\mu/\mpsi$ & $\torstar/\mpsi$ \\ \hline
 6.99  & 1/2 & 3 & .343(7) &  ---      & 1.79(4) & ---   \\
 7.78  & 1/4 & 6 & .370(7) & .139(23) & 1.90(4) & 4.8(8) \\
\hline
 6.99  & 1/2 & 4 & .221(6) &  ---      & 2.78(8) & ---     \\
 7.49  & 1/3 & 6 & .235(4) & .111(14) & 2.87(4) & 6.1(8) \\
 7.78  & 1/4 & 8 & .238(6) & .119(6)  & 2.85(6) & 5.6(3) \\
\hline
\end{tabular}
\end{center}
\caption{ Critical temperature $T_c$ in units of the torelon masses $\mu$
and $\torstar$, and the ratio of these to $\mpsi$.
Entries are grouped according to torelons of the same physical length
$L$, given in lattice units.
}
\label{tor_scaling}
\end{table}

\begin{table}
\begin{center}
\begin{tabular}{cccccccc}
\hline
$\bplc$ & $T_c$ & $L$ & $L$~[fm] & $\sigma a^2$ & $\sqrt{\sigma} L$ & 
                  $T_c/\sqrt{\sigma}$ & $\sqrt{\sigma}$~[MeV] \\ \hline
 6.99  & 1/2 & 3 & .99(2) & .603(1)  &  2.33(2)  & .644(5)  & 463(8)  \\
 7.78  & 1/4 & 6 & .98(3) & .142(2)   &  2.26(2)  & .664(5)  & 456(9)  \\
\hline
 6.99  & 1/2 & 4 & 1.37(2) & .631(15)  &  3.18(4)  & .630(7)  & 458(8)  \\
 7.49  & 1/3 & 6 & 1.28(2) & .266(4)   &  3.09(2)  & .647(5)  & 478(6)  \\
 7.78  & 1/4 & 8 & 1.29(3) & .148(4)   &  3.07(4)  & .651(8)  & 469(6) \\
\hline
\end{tabular}
\end{center}
\caption{String tension $\sigma$ defined by {\eq{sig_def}}, 
given in the dimensionless combinations with $T_c$ and $L$, 
and in MeV.  $\mpsi$ fixes the physical scales.  }  
\label{sigma_table}
\end{table}

Table~\ref{tor_table} lists the results of these fits for the ground 
state masses and several of the excited states.  For smaller lattices, 
we were unable to get accurate fits for the excited states, presumably 
due to the larger splitting, and settled for $2\times 2$ matrix fits to 
two hyperbolic cosines.  Our ground state masses $\mu$ fell below those 
of \rf{degrand}, from 3\% below at $T_c = 1/2$, increasing to $12\%$ below 
for $T_c = 1/4$.  This may indicate excited state
contamination in their fits, which would account for the increasing 
discrepancy with torelon length.  We did reproduce their result for the
torelon on a $3^3\times 16$ lattice using an unimproved Wilson action at
$\beta = 5.10$, obtaining 1.82(3), which we computed as a check.  In this 
case we were unable to reliably fit the excited state or include time steps 
much beyond the first couple.

In Table~\ref{tor_scaling} we present the critical temperature in units
of the torelon masses, as well as the ratio of these masses to $\mpsi$.  
The torelons we studied had two different physical lengths, determined
by the ratio of $L$ to $1/T_c$, and are grouped accordingly.  In most 
cases, these quantities are independent of $a$ to within statistical 
errors.  There is roughly a 5\% discrepancy for the coarsest lattice, 
which may indicate an $\order{a^4}$ effect.  \rf{degrand} found about 15\%
violations for an unimproved Wilson action, while results from their
improved (perfect) action scaled within errors.

We define the string tension $\sigma$ in terms of a torelon of finite 
length $L$ by
\be
 \sigma \equiv \mu/L + \pi/(3 L^2) \; ,
 \label{sig_def}
\ee
and demonstrate scaling in various dimensionless ratios involving $\sigma$
in Table~\ref{sigma_table}.  The extra Coulombic term in \eq{sig_def}
accounts for long wavelength fluctuations, and appears universally 
in quantized string models when $L$ is finite~\cite{string,torelon}.
Without it, we found that quantities in the last two columns for the 
two different physical lengths independently scale, but show a significant 
discrepancy when compared, providing confirmation for these models.  
Finally, using $\mpsi$ to determine the scale, we give values for the 
string tension in MeV.  Central values from these last two columns are 
$T_c/\sqrt{\sigma} = .649(5)$ and $\sqrt{\sigma} = 467(4)$~MeV.

\section{Conclusions}

We computed the critical temperature $T_c$ for the pure QCD deconfining
transition on coarse lattices, with spacing $a$ from .33~fm to .15~fm.  
By comparison with charmonium splittings and torelon masses measured
at these same spacings, we demonstrated that a gluon action improved to 
remove $\order{a^2}$ errors could measurements nearly free of lattice 
spacing errors.  In particular, that lattices with $N_s$ as small as 
three could give results for $T_c$ accurate to 5\%.  This precision is 
consistent with those found previously for charmonium and for the static 
quark potential.  We found central values for $T_c$ and the string 
tension $\sigma$ of $T_c = 300(6)~{\rm MeV}$, 
$T_c/\sqrt{\sigma} = .649(5)$ and $\sqrt{\sigma} = 467(4)$~MeV.

\section{Acknowledgments}

We performed many of the simulations on the IBM SP-2 supercomputer
at the Cornell Center for Theory and Simulation. 
This work was supported by grants from the NSF and DOE.

\end{document}